\documentclass[a4,twocolumn]{article}
\usepackage{graphicx}
\DeclareGraphicsExtensions{.pdf,.eps,.png,.jpg,.mps,.bmp}
\usepackage{algorithmic}
\usepackage{array}
\usepackage{color}
\usepackage{picture}
\usepackage{subfig}
\usepackage{amsmath,amssymb}
\usepackage{mathabx}
\usepackage{url}
\usepackage{epstopdf,times,amsmath,amssymb}

\newcommand{\argmin}{\mathop{\rm arg~min}\limits}

\title{Generating Similarity Map for COVID-19 Transmission Dynamics with Topological Autoencoder}
\author{Pitoyo Hartono \thanks{School of Engineering, Chukyo University, Nagoya, Japan}
}

\begin{document}
\maketitle

\section{Introduction}
At the end of 2019 China reported some cases of  pneumonia of unknown causes that rapidly became an outbreak \cite{tian2020}. On January 30 2020, The World Health Organization (WHO) declared the outbreak as a Public Health Emergency of International Concern, and on February 11 2020 the name for the coronavirus disease was announced as COVID-19. On March 11, COVID-19,  the human respiratory disease caused by the SARS-CoV2 virus, has been declared as a pandemic by the WHO with more than 100000 cases, in 114 countries as of that day.

Artificial Intelligence communities have been starting to work to tackle many aspects of COVID-19 crisis as nicely reviewed in \cite{bullock2020}. Data from this outbreak have already been compiled and made available for research communities \cite{Xu2020}.

In this short note, the author proposes the utilization of a topological neural network to map the global dynamics of COVID-19 spread. With so many countries affected by this crisis, and each country has different geographical, demographic and strategic aspects, hence, it is very difficult to obtain clear global view on how those differences are correlated with the failure or success of mitigating this crisis. 

The objective of this study is to offer a visualization method based on the time series data for COVID-19 patients numbers in many countries. While this visualization method does not directly offer insights for the means of crisis mitigation, it will offer an intuitive information on the similarities and differences between many countries in the transmission dynamics of this pandemic that will pontentially be  useful for further analysis and prediction.

\section{Topological Autoencoder}

\begin{figure*}
\centering
{\includegraphics[width=12 cm, height=7cm]{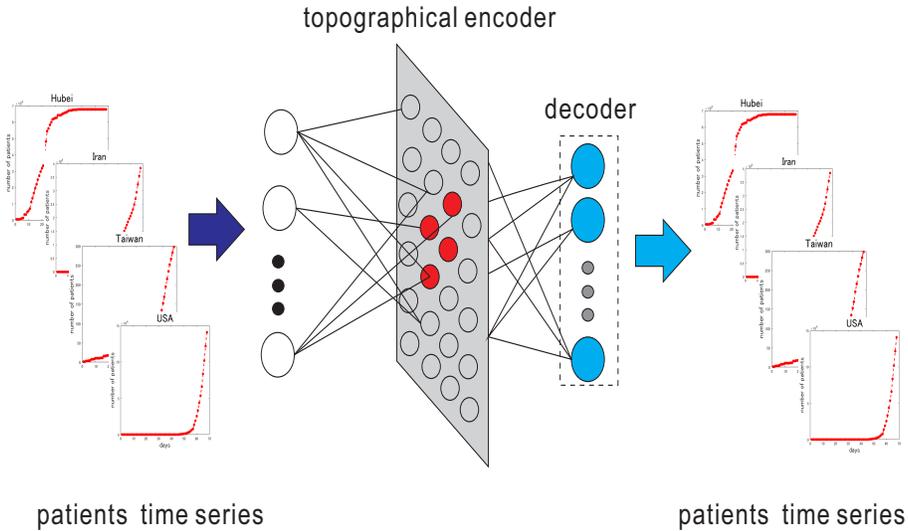}}
\caption{Topological Autoencoder}
\label{fig:outline}
\end{figure*}

The outline of Topological Autoencoder (TA) is shown in Fig. \ref{fig:outline}. TA is a hierarchical neural network having a low dimensional topological hidden layer that is a simplified version of Soft-supervised Topological Autoencoder (STA) proposed in \cite{hartono2019}. Originally, STA can be trained in supervised manner, unsupervised manner or the mix of both, and produces a two-dimensional topological map that allows human user to visualize the concept of the data. TA is a simplified version of STA in that it is only trained as an autoencoder in a unsupervised manner and produces a visible map that captures the inherent characteristics of the training data. 

In this preliminary study, TA is trained with time series of the numbers of COVID-19 patients in more than 240 countries or regions in the world that have been daily compiled by Center for Systems Science and Engineering (CSSE) at Johns Hopkins University \cite{dong2020} and the data are made available in a GitHub repository \cite{jhurepository2020}.

The hidden layer of TA is a topological map similar to Kohonen's Self Organizing Maps (SOM) \cite{Kohonen082}, \cite{Kohonen2013}, where hidden neurons are arranged in two-dimensional grid, and each neuron is associated with a reference vector that has the same dimensionality with the input.

The dynamics of TA is briefly explained as follows.

For a time series containing the daily numbers of COVID-19 cases in a certain country, $\textbf{X} \in \mathbb{R}^d$, TA selects the best matching unit, $win$  among all the reference vectors associated with the hidden units of STA as in Eq. \ref{eq:bmu}, where $\textbf{W}_j \in \mathbb{R}^d$ is the reference vector associated with the $j$-th hidden unit. Here, $d$ is the number of the days from the beginning of data collection on January 22 2020.

The output of the $j$-th hidden neuron, $H_j$ is shown in Eq. \ref{eq:hidden}, where $\sigma(j,win,t)$ is the neighborhood function defined in Eq.\ref{eq:temperature}. Here, $\sigma_{0} > \sigma_{\infty} > 0$ are the initial and final values of the annealing term, $t$ is the current epoch, while $t_{\infty}$ is the termination epoch.

\begin{eqnarray}
win &=& \argmin_j \|\textbf{X} - \textbf{W}_j \|   \label{eq:bmu} \\ 
H_j &=& \sigma(j,win,t) \mathrm{e}^{ -\frac{\| \textbf{X} - \textbf{W}_j\|^2}{\sigma^2}} \label{eq:hidden}
\end{eqnarray}

\begin{eqnarray}
\sigma(j,win,t) &=& exp(-dist(win,i)/S(t))  \label{eq:neighbor} \label{eq:temperature} \\
S(t) &=& \sigma_{\infty} + \frac{1}{2} (\sigma_0 - \sigma_{\infty}) (1+cos \frac{\pi t}{t_{\infty}}) \nonumber
\end{eqnarray}

The values of the $k$-th output neuron, $O_k $, is defined in Eq. \ref{eq:out} where $f(x)=\frac{1}{1+e^{-x}}$

\begin{equation}
O_k = f(({\textbf{V}_k})^t \textbf{H}) \nonumber  \\
\label{eq:out}
\end{equation}

Here, $\textbf{V}_k$ denotes the weight vector leading from the hidden layer to the $k$-th output neuron, $\textbf{H}=(H_1, H_2, \cdots, H_{N_{hid}})^t$ is the hidden layer output vector in which $N_{hid}$ is the number of hidden neurons.

The loss function is defined in Eq.  $\ref{eq:loss}$.

\begin{equation}
L = \frac{1}{2} \sum_{k}  (O_k - X_k )^2  \label{eq:loss}
\end{equation}

TA is then trained by minimizing the loss function with the standard stochastic gradient descent. The detail for the loss function derivations are explained in \cite{hartono2019}, \cite{hartono2015}.

The modifications of the reference vectors in the hidden layer of TA cause the formation of two-dimensional topological map, in which countries with similar transmission dynamics are assigned adjacently, while countries with dissimilar dynamics are separated by large space on the map. 

Utilizing this similarity map, we can gain intuitive understanding on global trend in COVID-19 transmission. The insights gained from visually analyzing this map are potentially helpful in designing strategy for controlling and thus mitigating the spread of this disease.

\section{Topological Maps for COVID-19 Transmission}

The following figures show a series of topological maps that reflect the dynamics of patients spread for around 240 countries or areas based on patients' numbers collected from January 22 2020 up to the date specified in the respective map. Each country is represented with a number according to the data numbering in John Hopkins University Repository. For visual clarity, the names of some countries or areas are shown on the map.

\begin{figure*}[htbp]
\begin{center}
\includegraphics[width=15 cm, height=15 cm]{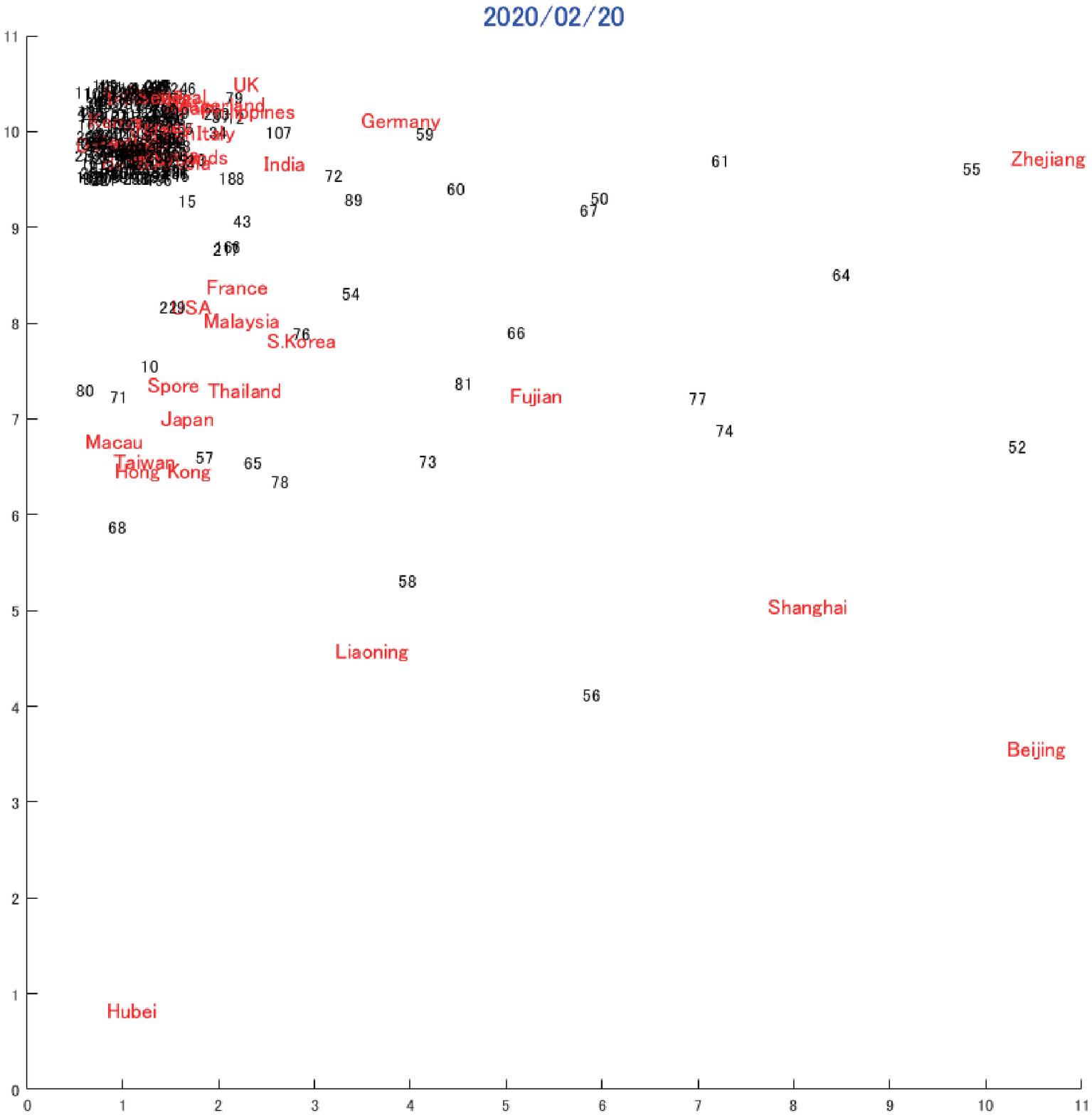}
\end{center}
\caption{Map Feb. 20 2020}
\label{fig:feb20}
\end{figure*}

Figure \ref{fig:feb20} shows the topological maps for patients spread up to February 20 2020, the early stage when COVID-19 patients were detected outside China. From this map, it can be observed that many countries occupy the upper left area on the map. The overlapping are due to the similar dynamics in the COVID-19 cases among many countries, in that most of them have identified very few patients at this time. At this time, Hubei, the epicenter for outbreak, was already an outlier on the map, as the number of patients in Hubei was in different magnitude compared to other areas. It can be also observed here that USA, France, South Korea, and  Asian countries as Malaysia, South Korea, Singapore, Thailand, Taiwan and Hong Kong were positioned adjacently, due to their similar dynamics, while many China's provinces as Zhejiang, Shanghai, Liaoning and naturally Hubei were outliers due to their early outbreaks.

\begin{figure*}
\begin{center}
\includegraphics[width=15 cm, height=15 cm]{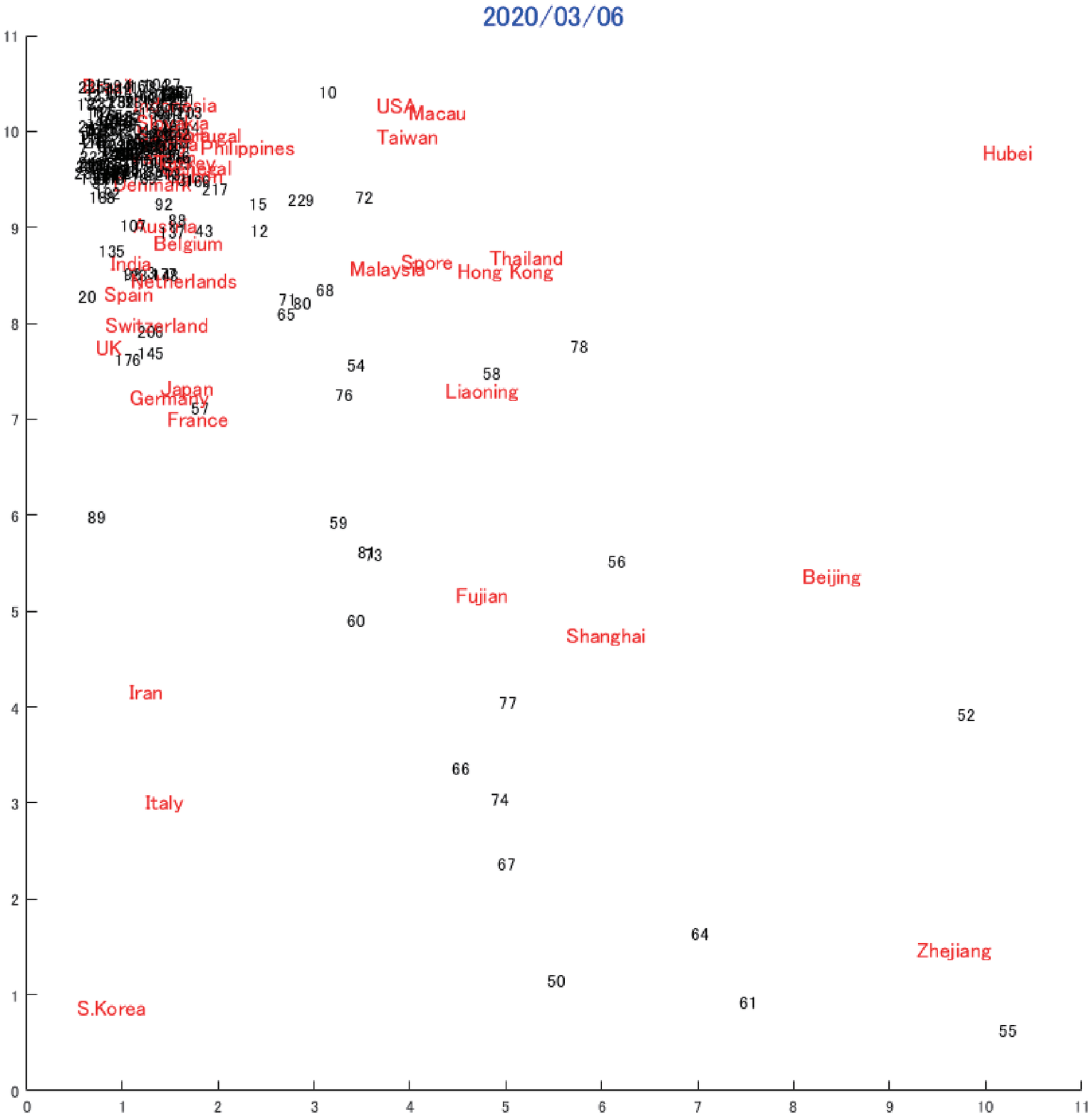}
\end{center}
\caption{Map March 6 2020}
\label{fig:march6}
\end{figure*}

Figure \ref{fig:march6} shows the map up to March 6. The upper left of the map was still crowded, but many countries can be seen to breakout from this overlapping area. This is due to many countries were beginning to develop different strategies resulting in different patients transmission dynamics. It is notable that Italy and Iran moved considerably from the crowded cluster due to the beginning of outbreaks in these two countries, while USA still shared the space, and thus has similar dynamics with Macau and Taiwan.

\begin{figure*}
\includegraphics[width=15 cm, height=15 cm]{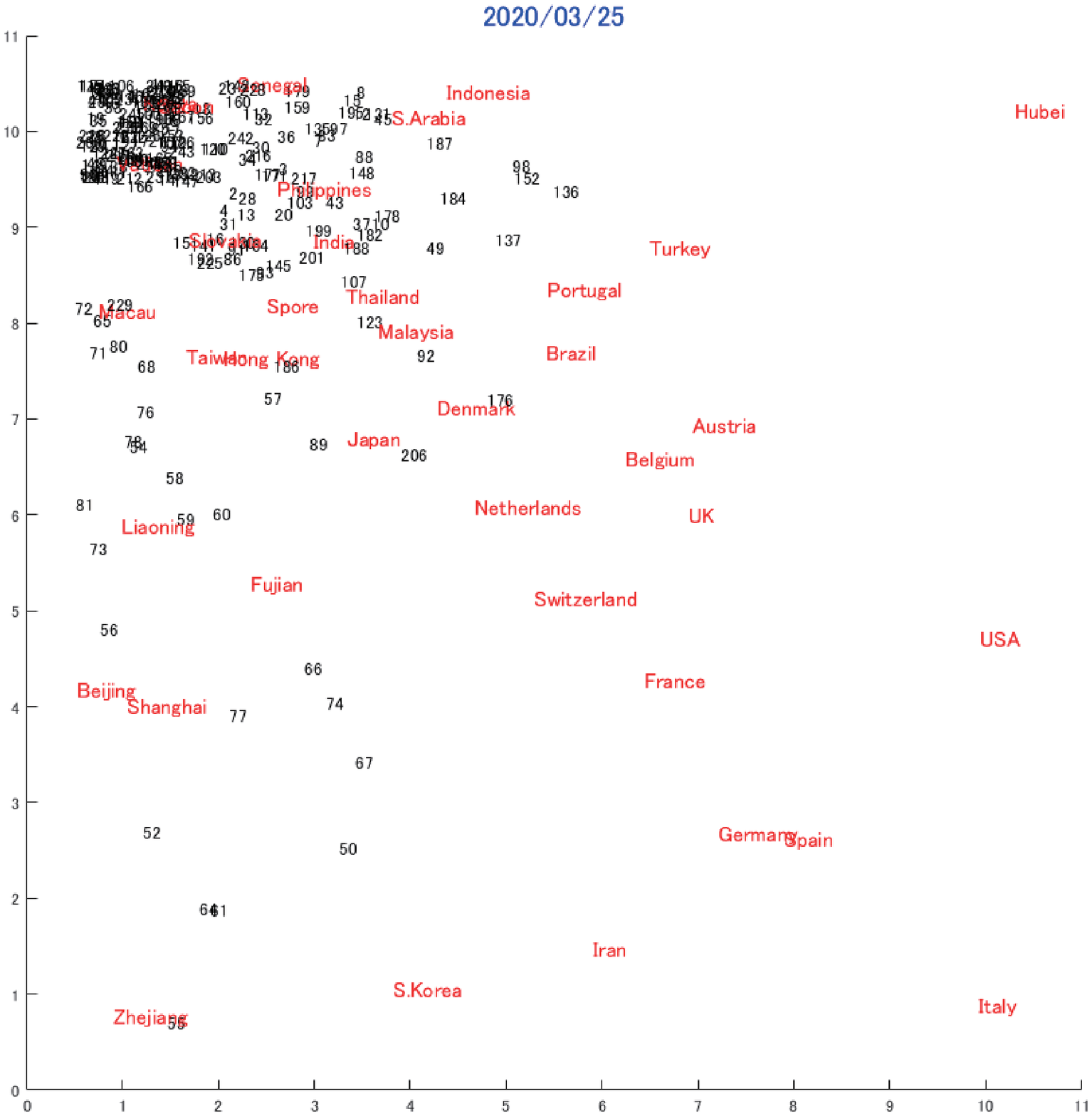}
\caption{Map March 25 2020}
\label{fig:march25}
\end{figure*}

Figure \ref{fig:march25} shows the map up to March 25, two weeks after World Health Organization (WHO)  declared COVID-19 as a pandemic. The left corner became less crowded as more and more countries developed their unique spread dynamics. Noticeably, South Korea moved away, due to rapid surge of virus transmissions in this period. Outbreaks occurred in USA, Spain and Italy as reflected by their positions on the map. Hubei, the original epicenter for this disease, many of Chinese's provinces and  South Korea kept their unique positions on the map because of their unique dynamics that saw early outbreak but than stabilized. 

\begin{figure*}
\includegraphics[width=15 cm, height=15 cm]{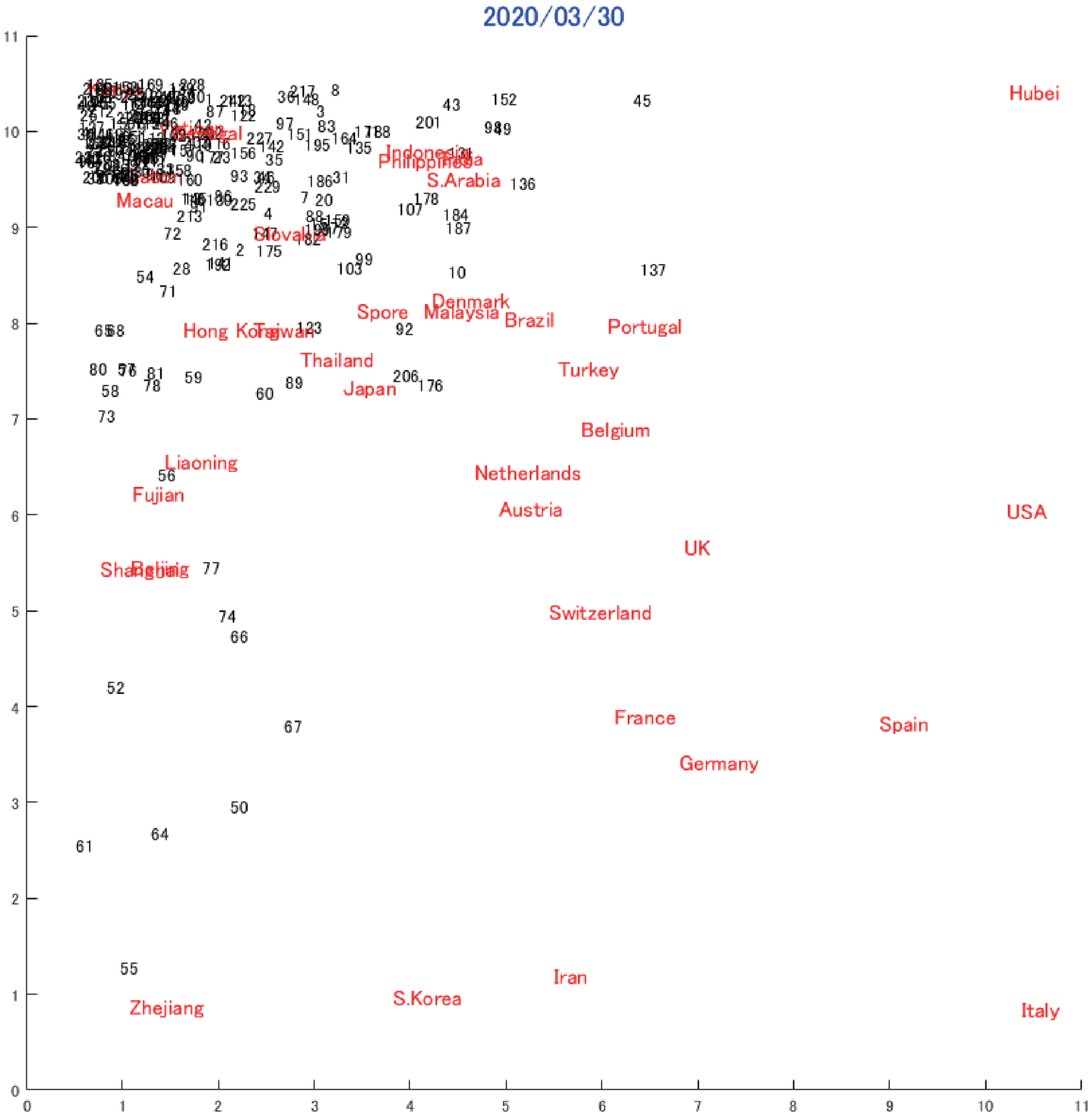}
\caption{Map March 30 2020}
\label{fig:march30}
\end{figure*}

Figure \ref{fig:march30} shows the map up to March 30, that did not significantly change from the previous map. This is because no noticeable new outbreaks, and hence many countries have stabilized their policies that results in the stabilization of their spread dynamics. The noticeable change is the split of Germany and Spain that previously shared a same dynamics. 

\begin{figure*}
\subfloat[Map Left: Patients]{\includegraphics[width=5 cm, height=5cm]{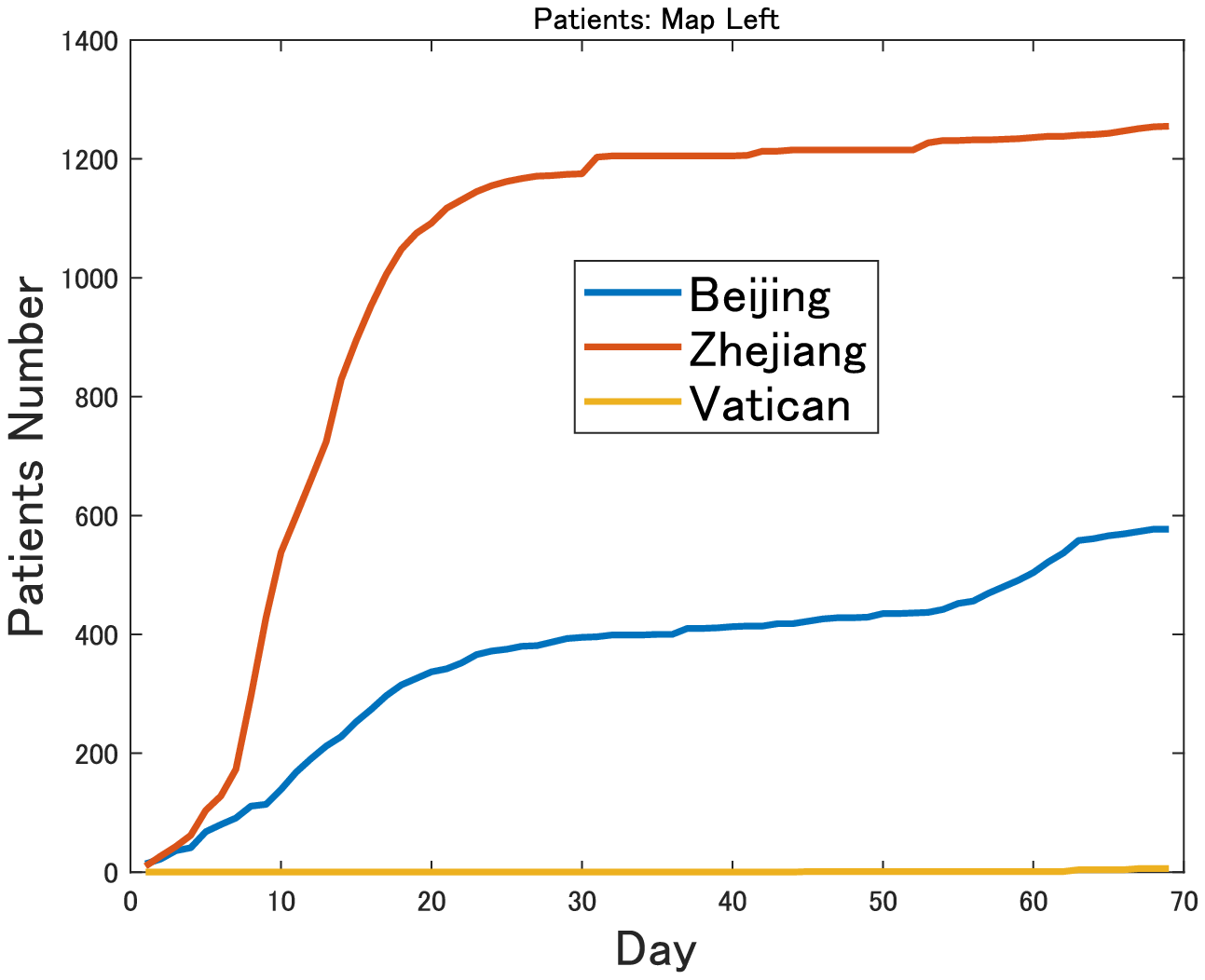}\label{fig:graphpatientleft}}
\subfloat[Map Mid: Patients]{\includegraphics[width=5 cm, height=5cm]{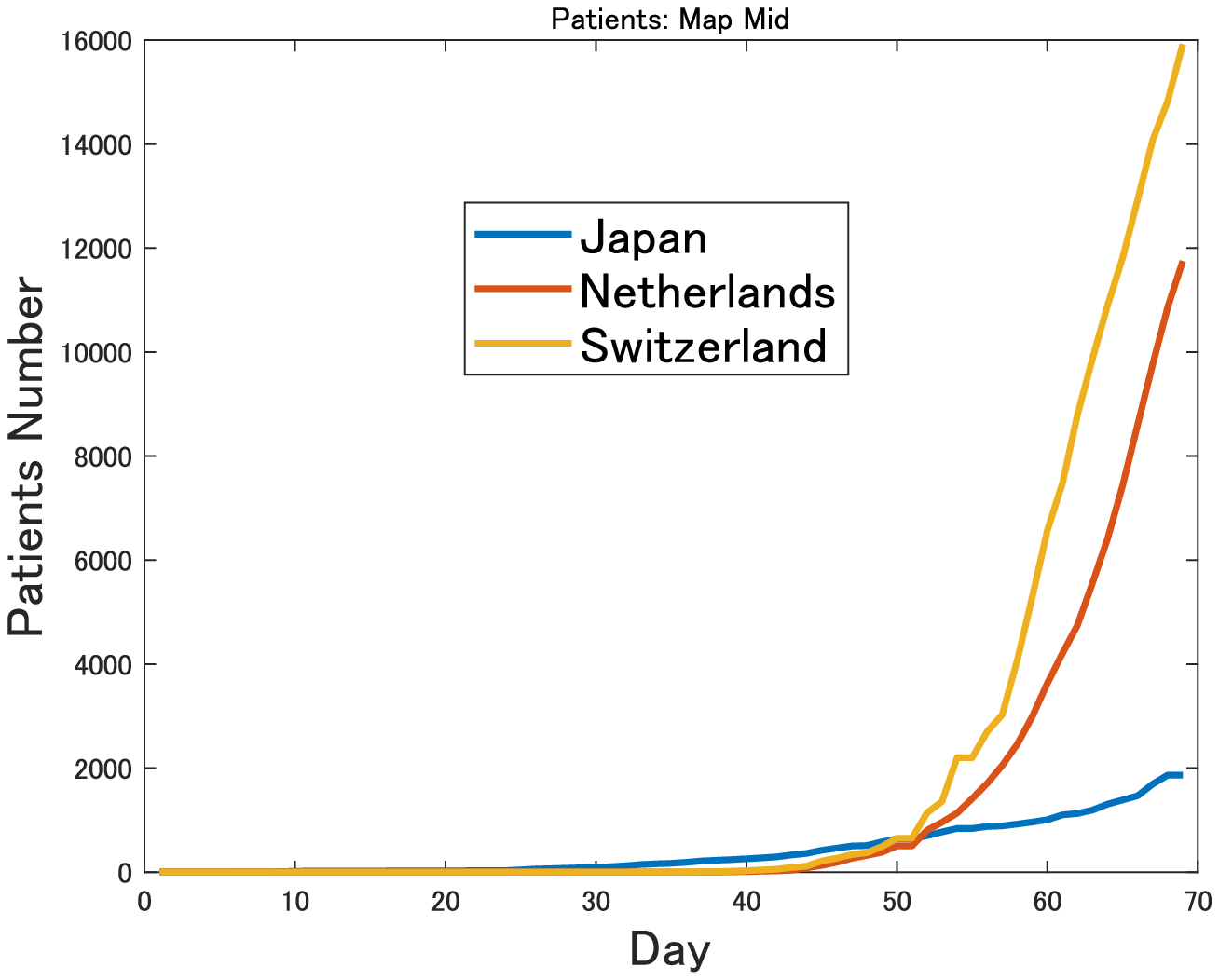}\label{fig:graphpatientmid}}
\subfloat[Map Right: Patients]{\includegraphics[width=5 cm, height=5cm]{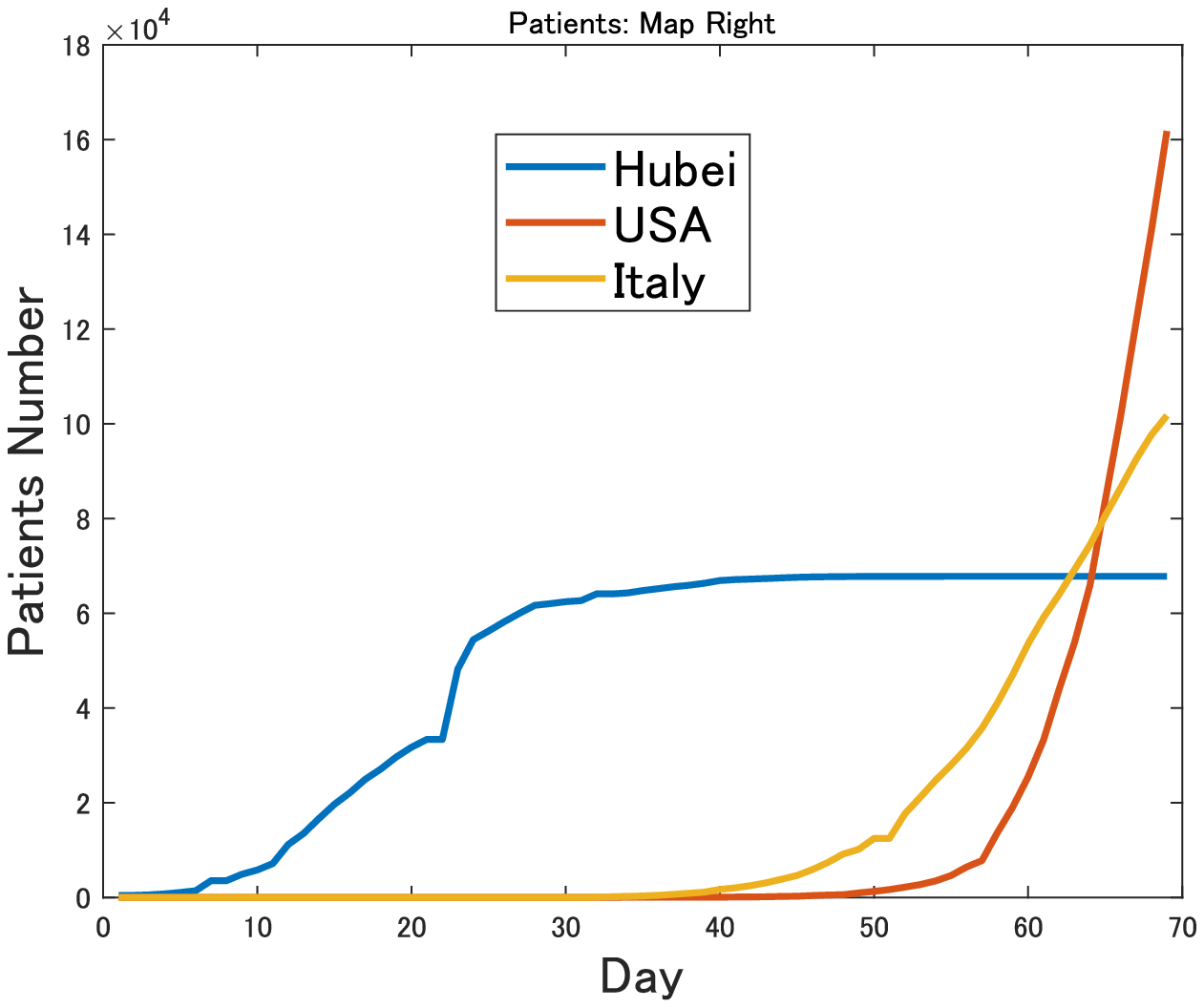}\label{fig:graphpatientright}}
\caption{Patients Spread}
\label{fig:patient}
\end{figure*}

Detailed analysis for this final map was executed to give better understandings on the topological characteristics of these maps. 

\begin{figure*}
\subfloat[Map Left]{\includegraphics[width=5 cm, height=5cm]{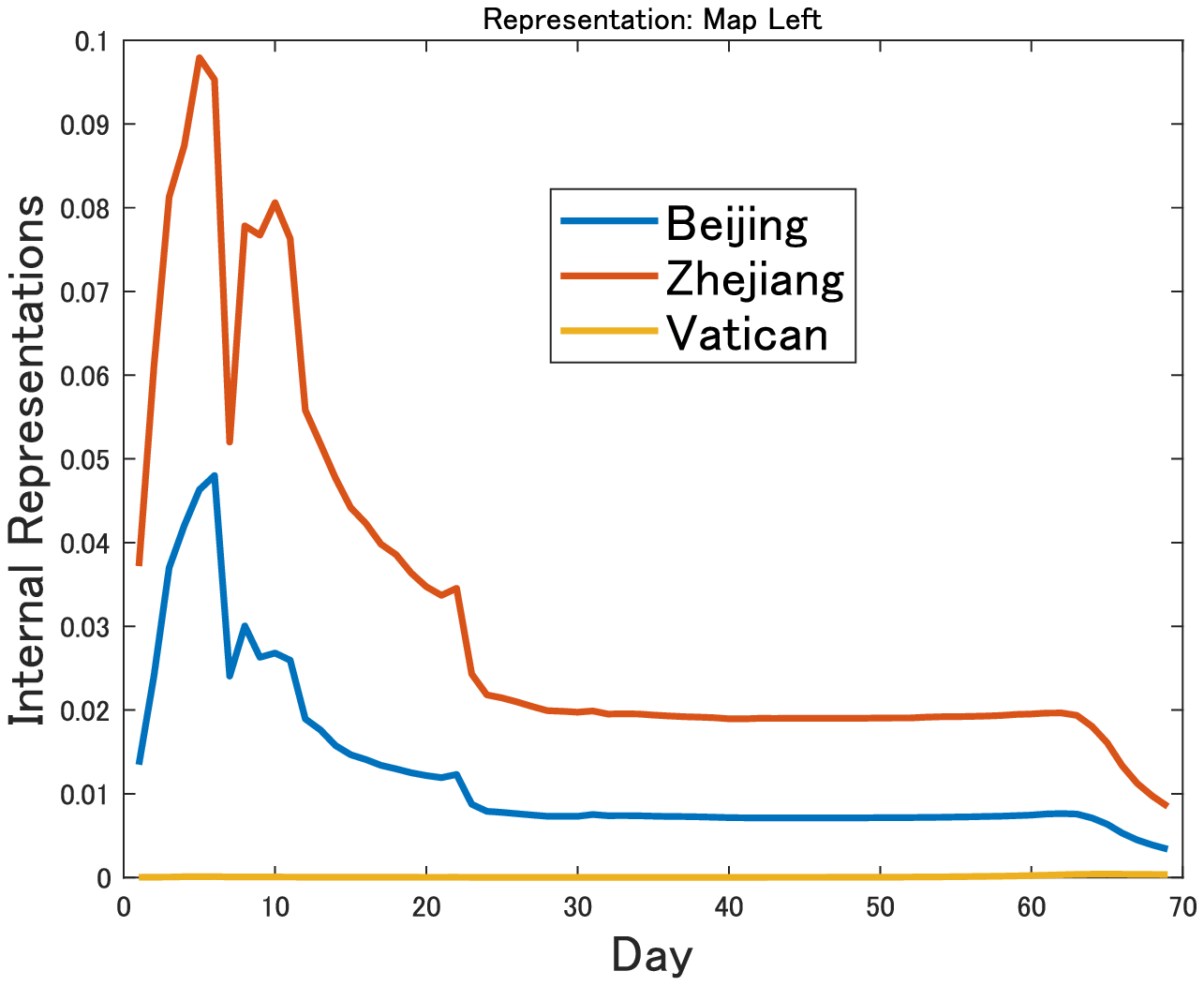}\label{fig:graphrepleft}}
\subfloat[Map Mid]{\includegraphics[width=5 cm, height=5cm]{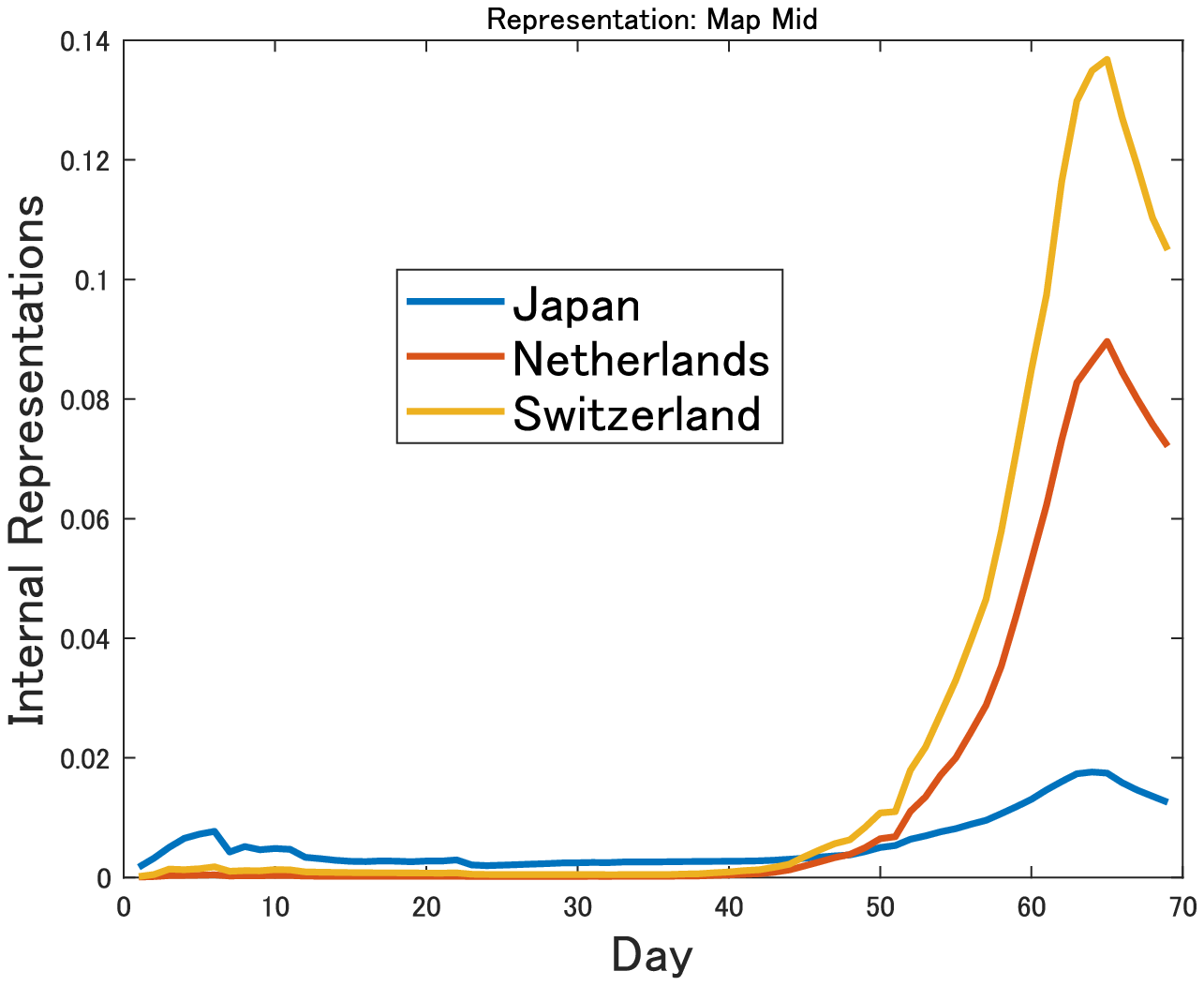}\label{fig:graphrepmid}}
\subfloat[Map Right]{\includegraphics[width=5 cm, height=5cm]{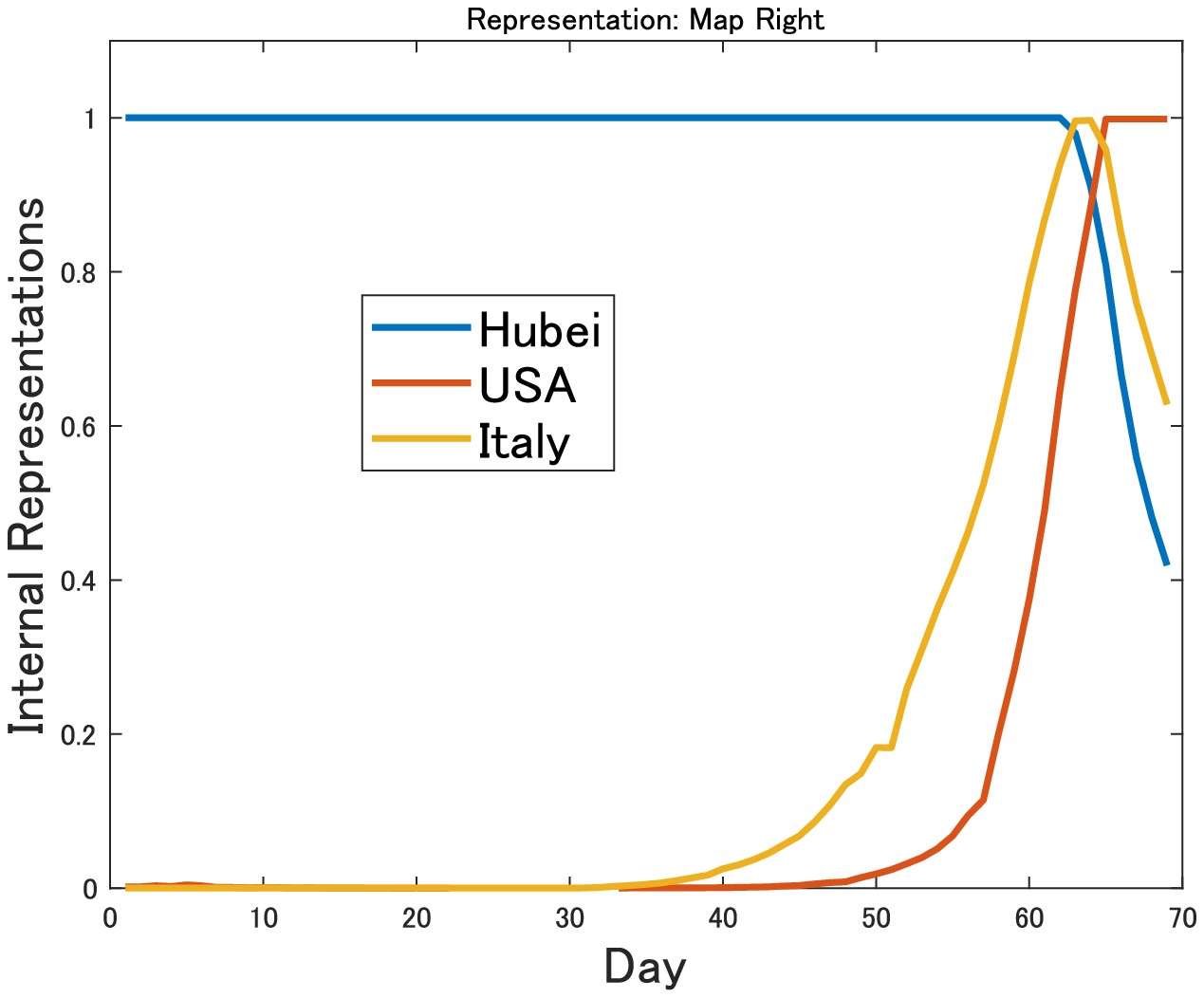}\label{fig:graphrepright}}
\caption{Internal Representations}
\label{fig:internalrep}
\end{figure*}

Figure \ref{fig:patient} shows the actual time series up to March 30 for some countries  divided based on the region they occupied on the map in Fig. \ref{fig:march30}. From the left area, as shown in Fig. \ref{fig:graphpatientleft}, Beijing, Zhejiang and Vatican had dissimilar dynamics that were reflected by their large distances on the map. Beijing and Zhejiang have both contained the outbreaks but had different dynamics and cumulative patient numbers, while Vatican had very few patients up to March 30. In the middle of the map Japan, Netherlands and Switzerland can be observed each with considerably different dynamics as shown in Fig. \ref{fig:graphpatientmid}. From the right side of the map, Hubei, USA and Italy as shown in Fig. \ref{fig:graphpatientright}. It is obvious that Hubei had stabilized while patients numbers were growing steeply in USA and Italy. Figure \ref{fig:internalrep} shows the internal representations, each one is the reference vector for the winning neuron associated with each country. Figure \ref{fig:graphrepleft} shows the representation for Beijing's, Zhejiang's and Vatican's dynamics, Fig. \ref{fig:graphrepmid} shows the representation for Japan's, Netherlands' and Switzerland's dynamics, while Fig. \ref{fig:graphpatientright} shows those of Hubei, USA and Italy. Here, it is important to notice that TA represents different dynamics with different internal representations that are important to recreate the original dynamics in the output layer of TA. Here, it is important to notice that some similar, although different dynamics may be represented by a single representation. The representation captures important features of those dynamics, and thus becomes a building block for them.

\begin{figure*}
\centering
\subfloat[Patients: S.Korea-UK]{\includegraphics[width=8 cm, height=4cm]{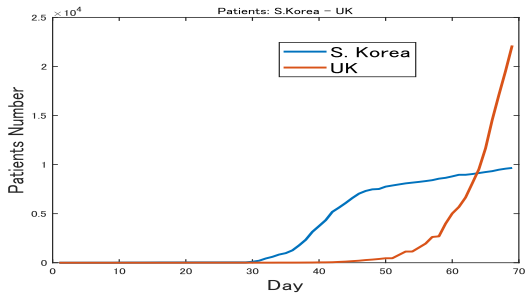}\label{fig:koreaukpatient}}
\subfloat[Representation: S.Korea-UK]{\includegraphics[width=8 cm, height=4cm]{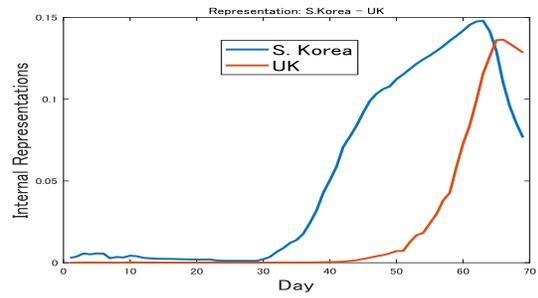}\label{fig:koreaukrep}}
\caption{S. Korea vs UK}
\end{figure*}

\begin{figure*}
\centering
\subfloat[Patients: Belgium-Netherlands]{\includegraphics[width=8 cm, height=4cm]{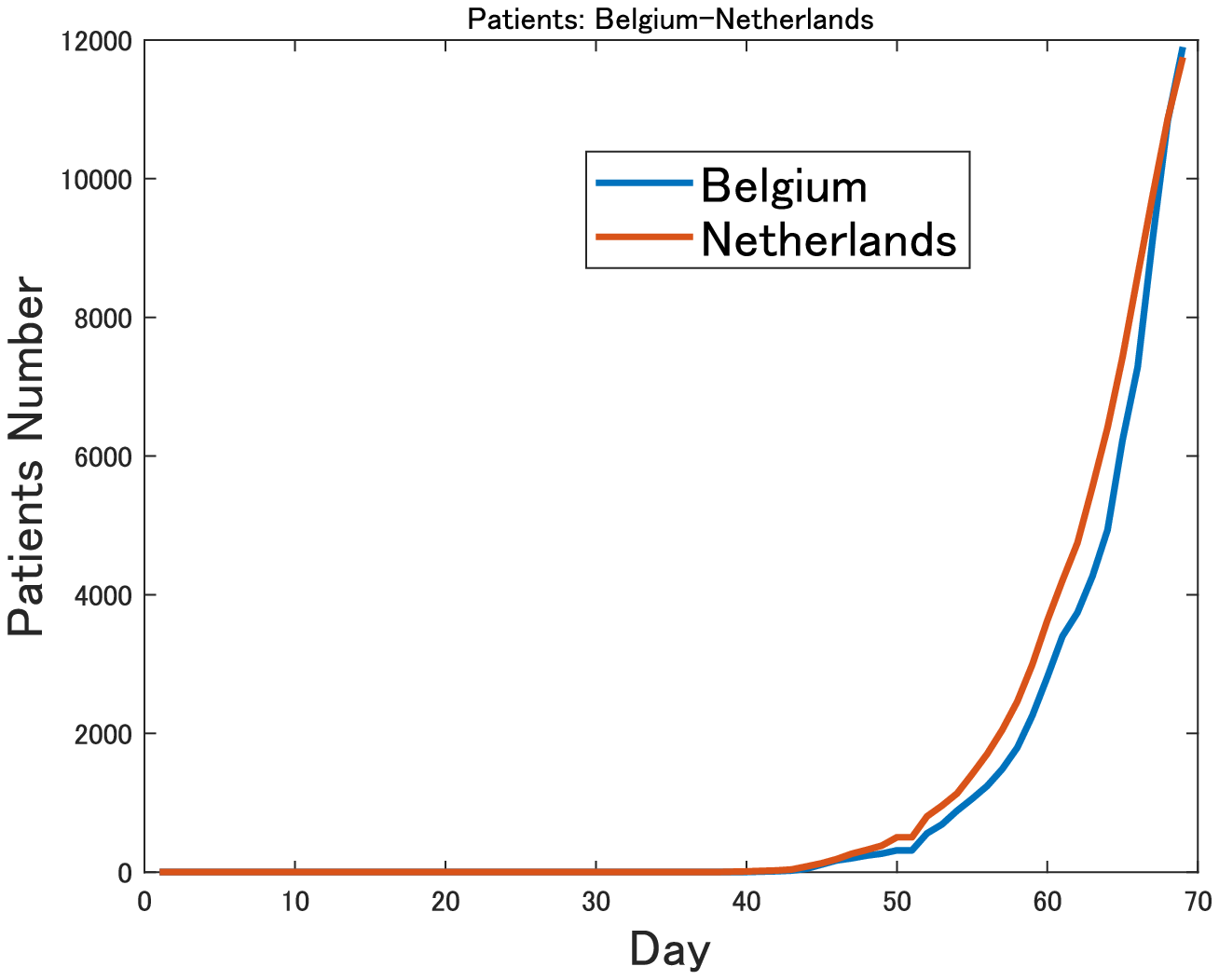}\label{fig:belnedpatient}}
\subfloat[Representation: Belgium-Netherlands]{\includegraphics[width=8 cm, height=4cm]{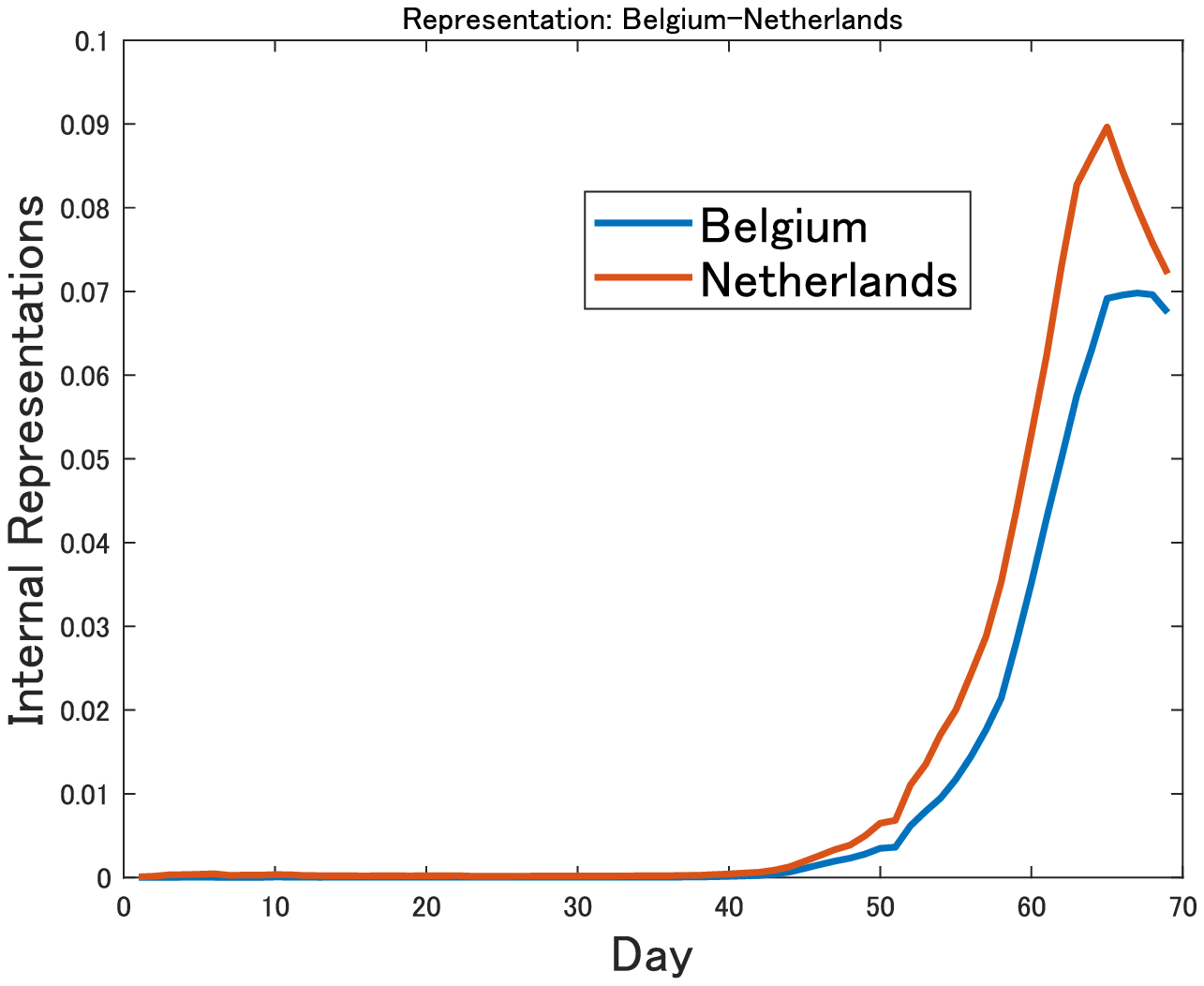}\label{fig:belnedrep}}
\caption{Belgium vs Netherlands}
\end{figure*}

To further explain the internal representation characteristics, the dynamics of South Korea and UK are shown in Fig. \ref{fig:koreaukrep}. These two countries are deliberately chosen because of their contrasting  transmission dynamics that are reflected by their largest distance on the map. South Korea are known for their success in containing the spread of this disease after an early outbreak, while UK, on March 30, was still struggling. Their internal representations are shown in Fig. \ref{fig:koreaukrep}, which clearly show the difference between these two countries. Here a clear example for the reflection of the differences in dynamics into the internal representations is shown.

Oppositely, Fig. \ref{fig:belnedpatient} shows the similar dynamics for Belgium and Netherlands that naturally are placed closely on the map in Fig. \ref{fig:march30}. The similarity in the disease spread dynamics is reflected in the similar representations in Fig. \ref{fig:belnedrep} and their positions on the map as well.

Through these preliminary examples, it can be empirically shown that the hidden representation of TA is able to capture the topological relations in different countries dynamics and thus can be utilized as an  analytical tool, for intuitively comparing the dynamics of many countries and potentially analyzing the correlation between those dynamics and strategies to mitigate this crisis.

\section{Conclusion}
In this unrefereed note, the author proposes the utilization of topological neural network for visualizing the global trend  for the  COVID-19 transmission. The initial experiments indicate that we can gain intuitive insights on the similarities and differences in the disease transmission among many countries. By visualizing the maps in time, we can also learn the shifts in each countries situations, which mirror their strategies. At this moment, this topological map is not directly useful for evaluating the efficiency of strategies executed by many countries to mitigate the crisis. However, the author believes that the intuitive visual understanding gained from these topological maps will be useful in comparing strategies among many countries, and thus for designing optimal strategies for mitigating this crisis. The direct usage of this map, is to cluster countries having similar dynamics, and train a time series predictor, for example Long Short-term Memory (LSTM) \cite{lstm}, for further predicting the future numbers of patients in those countries.

\bibliographystyle{hieeetr}
\bibliography{hartonobib} 

\end{document}